# Strongly localized polaritons in an array of trapped two-level atoms interacting with a light field

A.P. Alodjants, I.O. Barinov and S.M. Arakelian

Department of Physics and Applied Mathematics, Vladimir State University, Gorkiy str. 87, Vladimir, 600000, Russia

E-mail: alodjants@vlsu.ru

**Abstract.** We propose a new type of spatially periodic structure, i.e. polaritonic crystal (PolC), to observe a "slow"/"stopped" light phenomenon due to coupled atom-field states (polaritons) in a lattice. Under the tight-binding approximation, such a system realizes an array of weakly coupled trapped two-component atomic ensembles interacting with optical field in a tunnel-coupled one dimensional cavity array. We have shown that the phase transition to the superfluid Bardeen-Cooper-Schrieffer state, a so-called (BCS)-type state of low branch polaritons, occurs under the strong coupling condition. Such a transition results in the appearance of a macroscopic polarization of the atomic medium at non-zero frequency. The principal result is that the group velocity of polaritons depends essentially on the order parameter of the system, i.e. on the average photon number in the cavity array.

**Key words:** atomic polaritons, band-gap structures, slow light, tight-binding approximation, phase transition.

**PACS** 71.36.+c, 42.70.Qs, 42.50.Pq , 05.70.Fh

## 1. Introduction

Nowdays, the photonic band-gap micro- and nanostructures, i.e. photonic crystals, photonic crystal fibers, cavity arrays, etc., are known to be important tools for controlling light propagation properties in media, see e.g. [1]. Strong confinement of the photons and reduction of the group velocity of light due to the band-gap energy spectrum for one-, two- or three-dimensional boson-like systems plays an essential role in the solution of modern problems in both material science and information science. As an example, we mention here the application of 2D photonic band-gap structures in the process of creating high-density 3D optical memory devices, proposed, e.g. in [2].

The important features of photonic band-gap structures under discussion are connected with "slow" light, which is one of the promising fundamental physical phenomena that can be explored in the design of various quantum optical storage devices [3]. In particular, an effective reduction of the light group velocity is achieved in both hot and cold atomic ensembles [4-6], in resonant photonic band-gap structures, i.e. in coupled resonator optical waveguides (CROW) [7,8] and also, in different types of solid-state multilayer semiconductor systems [9]. The physical reason for observable reduction of the group velocity of light propagation in a medium is determined by the so-called dark and bright polaritons. The objects are bosonic quasi-particles representing a linear superposition of photon states in the external electromagnetic field and the macroscopic (coherent) excitations in the two-level oscillator system. The different methods applied to manipulate the group velocity for propagating optical pulses (like electromagnetically induced transparency, spin echoes, etc.) represent a necessary *polaritonic toolset* to achieve the long-time coherence for the coupled atom-light system, cf. [10]. The fact is necessary for storing the quantum optical information.

In our previous papers [11,12] we proposed the use of cavity polaritons arising due to the matter-field interaction under the strong coupling condition for cloning and spatial storing of quantum optical information. At the same time, we showed that the polaritonic model for the atom-field interaction is valid within the low excitation density limit when the atoms populate predominantly for their ground state.

The low branch polaritons under discussion can form Bose-Einstein condensation (BEC) due to phase transition occurring in the cavity [12]. In fact, evidence for BEC for 2D-gas exciton-polaritons in semiconductor (Cd-Te) microstructures at the temperature of 5K has been reported recently [13]. However, the achieved lifetime of such a coherent excitation state is determined by the *picosecond* time scale, and it seems to be short enough to make a quantum memory device practically useful.

In the atomic system, the lifetime of polaritons is limited by spontaneous emission of radiation for atoms, and can typically be at the *nano-scale*, cf.[14]. However, the activity in the area of laser field manipulation of a large number of ultra-cold atomic ensembles being under the BEC conditions [15] paves the way to further investigation of "slow" light phenomenon in coupled atom-light systems. For that case, the macroscopic array of atomic condensates is produced by trapping, cooling and localization of ultra-cold atoms in one- or two- dimensional optical lattices and gives an opportunity to study various aspects of the physics of phase transitions.

On the other hand, modern nanofabrication and nanophotonic technologies make it possible to study slow light and phase transitions of polaritons using the array of photonic cavities (or coupled resonator waveguides) doped by two-level atoms, see e.g. [16-18]. Technologically, such artificial structures can be produced with the help of photonic crystal structures doped by two-level atoms, cf. [1].

In this respect, we have recently proposed spatially periodical atomic structures, so-called *polaritonic crystal* (PolC), being a lattice of weakly coupled two-component trapped atomic ensembles interacting with the optical field in the cavity [19]. A remarkable feature of such structures is the possibility of complete polariton localization by analogy with light localization in photonic crystals (see e.g. [1]) as well as with exciton localization in quasiperiodic structures in solid-state physics, cf.[20]. In particular, we discuss here the behavior of the so-called phonon polaritons propagating in photonic crystals and/or in semiconductor structures developed in the framework of terahertz polaritonics, see e.g. [10, 21].

The principal goal of this paper is further development of atomic polariton theory with respect to observing the group velocity reduction for light being propagated under the second-order phase transition in coupled atom-field states in the lattice. We examine the influence of excitation density and photon number (order parameter) on control of the polariton group velocity in the PolC system with a band-gap structure.

The paper is arranged as follows. In section 2, we develop the quantum field theory approach for the model of a polaritonic band-gap structure with facilities for trapping and manipulating atoms by laser field. We focus our attention on the main properties of polaritons within the low excitation density limit in a spatially periodic atomic structure. In section 3, we consider a thermodynamic approach to study polariton properties in the PolC structure for practical (high enough) temperatures and fixed excitation densities. We have shown that the phase transition in such an atomic system is very close to the normal metal-superconductor transition established by the Bardeen-Cooper-Schrieffer (BCS) theory, see e.g. [22]. Finally, we examine a polariton group velocity depending on the temperature in the presence of the normal state of the system, i.e. the superfluid phase transition. In conclusion, we summarize the obtained results.

**2. Polaritons in the atomic array: the model of polaritonic crystal (PolC)**

A general model for a 2D-polaritonic crystal is presented in Fig.1. We propose an array of $M$ single-mode microcavities with the nearest-neighbor interactions in the XY-plane. Each of the cavities contains an ensemble of ultracold two-level atoms with $|a\rangle$ and $|b\rangle$ internal levels interacting with quantized electromagnetic fields in the $z$-direction. There exist different ways to produce such a system experimentally.

One of these is based on the confinement of two-level atoms in a special photonic band-gap structure, i.e. photonic crystal or CROW [7]. But in contrast with the papers [16,17], where coupled resonators with doped atoms are studied, in our configuration (see Fig.1) the overlapping of optical fields and atomic wave functions take place in the XY-plane being *orthogonal* to the cavities axis $z$.

Another promising candidate for experimental realization of such a system is connected with trapping atoms in hollow core optical fibers (photonic crystal or microstructure), see e.g. [1,23]. In this case, we have to describe a polaritonic waveguide array representing some extension for a tunnel-coupled optical fiber and/or for waveguide array, which is well known in nonlinear optics, cf. [24,25].

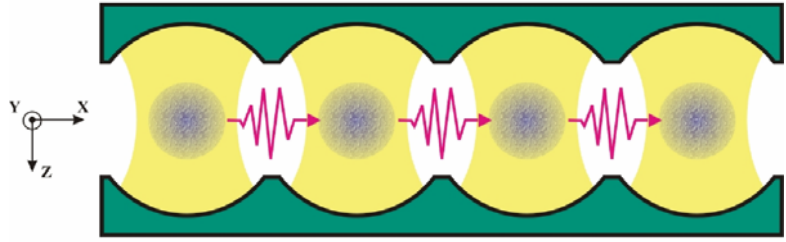

**Fig.1**. Model of polaritonic crystal: the array of the microcavities with macroscopically small number of two-level atoms.

The total Hamiltonian $H$ for the atom-light coupled system in Fig.1 can be represented as:
$$H = H_{at} + H_{ph} + H_{\text{int}}, \tag{1}$$
where $H_{at}$ is a part of the Hamiltonian that describes two-level atoms in a periodical structure, the term $H_{ph}$ is responsible for the photonic field distribution and the term $H_{\text{int}}$ characterizes the atom-light interaction in each cavity. In some approximations, Hamiltonian $H$ can be written down as (for more details see also [19]))

$$H_{at} = \hbar \sum_{n=1}^{M} \left( \omega_{n,at}^{(a)} a_n^\dagger a_n + \omega_{n,at}^{(b)} b_n^\dagger b_n - \frac{\gamma_a}{2}\left(a_n^\dagger a_{n-1} + a_n^\dagger a_{n+1} + H.C.\right) - \frac{\gamma_b}{2}\left(b_n^\dagger b_{n-1} + b_n^\dagger b_{n+1} + H.C.\right) \right), \tag{2a}$$

$$H_{ph} = \hbar \sum_{n=1}^{M} \left( \omega_{n,ph} \psi_n^\dagger \psi_n - \frac{\alpha}{2}\left(\psi_n^\dagger \psi_{n-1} + \psi_n^\dagger \psi_{n+1} + H.C.\right) \right), \tag{2b}$$

$$H_{\text{int}} = \hbar \sum_{n=1}^{M} \frac{g_n}{\sqrt{N}} \left( \psi_n^\dagger a_n^\dagger b_n + b_n^\dagger a_n \psi_n \right), \tag{2c}$$

where the annihilation (creation) operators $a_n$ ($a_n^\dagger$), $b_n$ ($b_n^\dagger$) in Eq.(2a) characterize the dynamical properties of atomic ensembles (atomic quantum modes) at the lower ($|a\rangle$) and upper ($|b\rangle$) levels in the $n$-site array ($n=1,2,....,M$) respectively (the $\hbar\omega_{n,at}^{(a)}$ and $\hbar\omega_{n,at}^{(b)}$ characterize the energy of atoms at the levels). The coupling coefficients $\gamma_{a,b}$ are the nearest-neighbor hopping constants that depend on the atomic cloud wavefunction overlapping integrals. These wavefunctions are real (Wannier) functions responsible for spatial distribution of ultracold $n$-site atoms under the so-called tight-binding approximation, cf. [26].

The annihilation (creation) operators $\psi_n$ ($\psi_n^\dagger$) in Eq.(2b) describe the temporal behavior of a single photonic mode with angular frequency $\omega_{n,ph}$ located at the $n$th cavity. Parameter $\alpha$ characterizes a spatial field overlapping between the neighbor cavities.

The interaction of two-level atoms with the quantized electromagnetic field in Eq.(2c) is considered under the rotating wave approximation and determined by the constant $g_n$. For simplicity, we assume all cavities being identical to each other and contain the same atom number $N_n = N$. At the same time, we suppose that the atom-light coupling coefficients in Eq.(2c) are equal to each other for all sites, i.e. $g \equiv g_1 = ... = g_M$. In addition, we consider some reasonable approximations for the problem.

First, we propose that atomic gas be an ideal one consisting of non-interacting particles.

Second, we assume that spatial degrees of freedom are "frozen" for both atomic and optical systems in the cavities. For the case, the characteristic angular frequencies $\omega_{n,at}^{(a,b)}$ ($\omega_{n,ph}$) also depend on the trapping potential of the atoms (photons) for each site. This approach is valid for a relatively small number of atoms, i.e. when $N \leq 10^4$, cf. [27].

Third, we are working under the so-called strong atom-light coupling regime for which the coupling parameter $g$ for each site of the lattice is much larger than the inverse coherence time $\tau_{coh}$ of combined atom-optical system, i.e. (cf. [11,12])

$$g \gg 2\pi/\tau_{coh}. \tag{3}$$

Physically, $\tau_{coh}$ is the necessary time to achieve a thermal equilibrium for the atomic system under the interaction with the quantized optical field in the PolC structure. We also neglect spectral line broadening mechanisms in accordance with condition (3) at low temperatures (mK). As a result, we have a pure quantum (or thermodynamically equilibrium) states for the coupled atom-light system.

Let us proceed to the $\vec{k}$-representation in expressions (1) and (2). Taking into account the periodical properties of the system we can represent the operators $\psi_n$, $a_n$, $b_n$ in the form:

$$a_n = \frac{1}{\sqrt{M}}\sum_k a_{\vec{k}} e^{i\vec{k}\vec{n}}, \quad b_n = \frac{1}{\sqrt{M}}\sum_k b_{\vec{k}} e^{i\vec{k}\vec{n}}, \quad \psi_n = \frac{1}{\sqrt{M}}\sum_k \psi_{\vec{k}} e^{i\vec{k}\vec{n}}, \tag{4}$$

where $\vec{n}$ is the lattice vector.

To be more specific, we consider below a *one dimensional* spatially periodical structure for PolC, i.e. $\vec{k}\vec{n} = nk_x\ell$, $n=1,2,...,M$, where $\ell$ is a lattice constant. Substituting Eq.(4) into Eq.(1) and taking into account the relation $\frac{1}{M}\sum_n e^{i(\vec{k}-\vec{k}')\vec{n}} = \delta_{\vec{k}\vec{k}'}$, we arrive at the $\vec{k}$-space expression for Hamiltonian (2), i.e.

$$H = \hbar\sum_{\vec{k}}\left(\omega_{ph}(k)\psi_{\vec{k}}^\dagger\psi_{\vec{k}} + \frac{\omega_{at}(k)}{2}\left(b_{\vec{k}}^\dagger b_{\vec{k}} - a_{\vec{k}}^\dagger a_{\vec{k}}\right) + \frac{g}{\sqrt{N_{tot}}}\sum_{\vec{q}}\left(\psi_{\vec{k}}^\dagger a_{\vec{q}}^\dagger b_{\vec{k}+\vec{q}} + b_{\vec{k}+\vec{q}}^\dagger a_{\vec{q}}\psi_{\vec{k}}\right)\right), \tag{5}$$

where $N_{tot} = NM$ is the total number of atoms for all sites.

In Eq.(5), the angular frequencies $\omega_{ph}(k)$ and $\omega_{at}(k)$ characterize the dispersion properties of the photonic and atomic system in the PolC structure, respectively; the parameters are determined by:

$$\omega_{ph}(k) = \omega_{n,ph} - 2\alpha\cos k\ell, \quad \omega_{at}(k) = \omega_{n,at}^{(b)} - \omega_{n,at}^{(a)} - 2\gamma\cos k\ell, \tag{6a,b}$$

where we introduce the effective coupling coefficient $\gamma = \gamma_b - \gamma_a$ for atoms in the lattice.

The Hamiltonian (5) has a diagonal form

$$H = \hbar \sum_{\vec{k}} \Omega_1(k) \Xi_{1,\vec{k}}^\dagger \Xi_{1,\vec{k}} + \hbar \sum_{\vec{k}} \Omega_2(k) \Xi_{2,\vec{k}}^\dagger \Xi_{2,\vec{k}}, \qquad (7)$$

when we use Bogoliubov transformations:

$$\Xi_{1,\vec{k}} = \vartheta_1 \psi_{\vec{k}} + \vartheta_2 P_{\vec{k}}, \quad \Xi_{2,\vec{k}} = \vartheta_1 P_{\vec{k}} - \vartheta_2 \psi_{\vec{k}}. \qquad (8)$$

The annihilation operators $\Xi_{1,\vec{k}}$, $\Xi_{2,\vec{k}}$ in Eq.(8) characterize two types of quasiparticles due to the atom-field interaction, i.e. upper and lower branch polaritons, respectively, propagating in the $x$-direction of the periodical structure in Fig.1. In Eq.(8) $P_{\vec{k}} = \frac{1}{\sqrt{N_{tot}}} \sum_{\vec{q}} a_{\vec{q}}^\dagger b_{\vec{k}+\vec{q}}$ is an operator of collective polarization for atoms with quasi-momentum $\vec{k}$; the quantities $\vartheta_{1,2}$ are Hopfield coefficients:

$$\vartheta_{1,2}^2 = \frac{1}{2}\left(1 \pm \frac{\delta}{\sqrt{\delta^2 + 4g^2}}\right). \qquad (9)$$

The parameters satisfy the normalization condition $\vartheta_1^2 + \vartheta_2^2 = 1$.

Here we focus on the low excitation density limit when all atoms in the cavities occupy mostly the ground level $|a\rangle$, i.e. $\sum_{\vec{q}} \langle b_{\vec{q}}^\dagger b_{\vec{q}} \rangle \ll \sum_{\vec{q}} \langle a_{\vec{q}}^\dagger a_{\vec{q}} \rangle \simeq N_{tot}$, cf. [11]. In the case, both operators $P_{\vec{k}}$ and $P_{\vec{k}}^\dagger$ for atomic polarization and polariton operators $\Xi_{1,\vec{k}}$, $\Xi_{2,\vec{k}}$ satisfy bosonic commutation relations

$$\left[P_{\vec{k}}; P_{\vec{k}}^\dagger\right] = \frac{1}{N_{tot}} \sum_{\vec{q}} \left(a_{\vec{q}}^\dagger a_{\vec{q}} - b_{\vec{k}+\vec{q}}^\dagger b_{\vec{k}+\vec{q}}\right) \simeq 1, \qquad (10a)$$

$$\left[\Xi_{i,\vec{k}}; \Xi_{i,\vec{q}}^\dagger\right] \simeq \delta_{ij} \delta_{\vec{k}\vec{q}}. \qquad (10b)$$

The characteristic angular frequencies $\Omega_{1,2}(k)$ in Eq.(7) determine the dispersion relation for polaritons in a band-gap structure, and are defined as follows:

$$\Omega_{1,2}(k) = \frac{1}{2}\left[\omega_{at}(k) + \omega_{ph}(k) \pm \omega_R(k)\right], \qquad (11)$$

where $\omega_R(k) = \left(\delta^2 + 4g^2\right)^{1/2}$ is the Rabi splitting angular frequency, $\delta = \omega_{ph}(k) - \omega_{at}(k) = \tilde{\delta} - 2(\alpha - \gamma)\cos k\ell$ is the phase mismatch depending on quasi-momentum $k$, $\tilde{\delta} \equiv \omega_{n,ph} - (\omega_{n,at}^{(b)} - \omega_{n,at}^{(a)})$ is the detuning for $k\ell = \pi/2 + \pi n$, $n = 0,1,...$.

By using Eqs. (10) for the masses of the upper ($m_1$) and lower ($m_2$) polariton branch one can obtain:

$$m_{1,2} \equiv \hbar \left(\frac{\partial^2 \Omega_{1,2}(k)}{\partial^2 k}\bigg|_{k=0}\right)^{-1} = \frac{2 m_{at} m_{ph} \left(\tilde{\Delta}^2 + 4g^2\right)^{1/2}}{(m_{at} + m_{ph})\left(\tilde{\Delta}^2 + 4g^2\right)^{1/2} \pm (m_{at} - m_{ph})\tilde{\Delta}}, \qquad (12)$$

where we introduce a maximal detuning $\tilde{\Delta} = \tilde{\delta} - 2(\alpha - \gamma)$ accessible for $k = 0$; $m_{at} = \frac{\hbar}{2\gamma\ell^2}$ and $m_{ph} = \frac{\hbar}{2\alpha\ell^2}$ are the effective masses of atoms and photons in the lattice. In a real experimental situation (when relation $m_{ph}/m_{at} \ll 1$ (i.e. $\alpha \gg \gamma$) takes place), the masses of polaritons $m_{1,2} = \dfrac{2m_{ph}}{1 \pm m_{ph}/m_{at}} \simeq 2m_{ph}$ are small enough for resonant atom-optical interaction for $\tilde{\Delta} = 0$.

To estimate the expected effects, we consider that a quantized optical field interacts with ensembles of two-level rubidium atoms having a mean resonance frequency of transition $382 THz$ and the wave vector of resonant light field $k_z \approx 0.8 \times 10^7 m^{-1}$ ($\lambda \approx 785 nm$), corresponding to the weighed mean of the rubidium D-lines. In this case, the effective photon mass can be estimated as $m_{ph} \approx 2.8 \times 10^{-36} kg$ which results in the low branch polariton mass $m_2 \simeq 5.6 \times 10^{-36} kg$. We also suppose a zero detuning Rabi splitting frequency (the atom-field coupling parameter) to be $g/2\pi = 500 MHz$ for all numerical simulations made below.

In Fig.2 the reduced dispersion relation $\tilde{\Omega}_{1,2}(k) \equiv \left[\Omega_{1,2}(k) - (\omega_{n,at}^{(b)} - \omega_{n,at}^{(a)})\right]/2g$ for the upper and lower branch of polaritons is presented in the first Brillouin zone of the periodical structure. The spontaneous emission lifetime for rubidium D-lines is $\tau_{spont} \approx 27 ns$; the value corresponds to spontaneous emission rate frequency being equal to about $37 MHz$. Condition (3) is still fulfilled in the case under discussion.

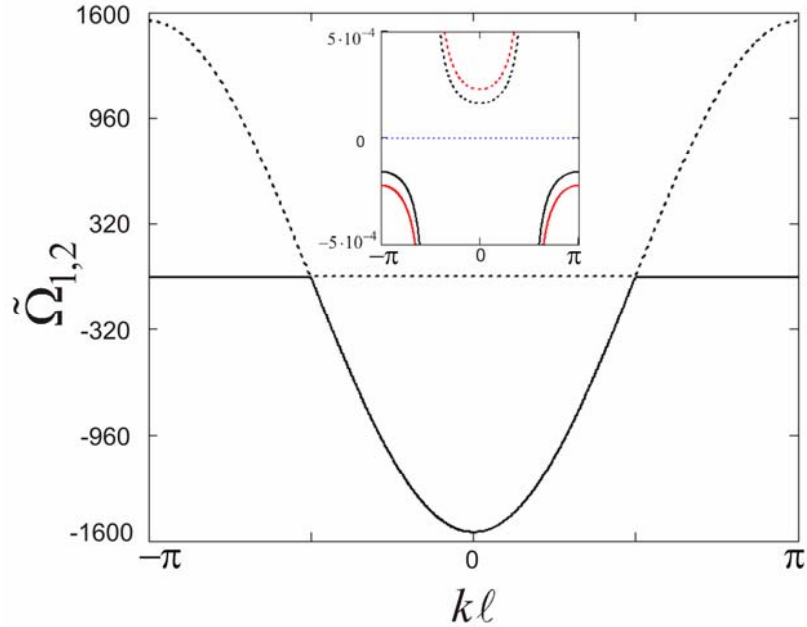

**Fig.2**. Dependences of characteristic angular frequencies for the upper ($\tilde{\Omega}_1(k)$) and lower ($\tilde{\Omega}_2(k)$) branches of polaritons on the reduced quasi-momentum (i.e. Bloch vector) $k\ell$ in the first Brillouin zone. The angular frequency detuning is $\tilde{\delta} = 0$, and the lattice constant is $\ell = 2\mu m$. In the inset, the frequency gap between two branches is presented. The red curves correspond to polariton dispersion taking into account cavity field amplitudes $\lambda_{\vec{k}} = \left(\langle \psi_{\vec{k}}^\dagger \psi_{\vec{k}} \rangle / N_{tot}\right)^{1/2} = 0.5$ and excitation density $\rho_{\vec{k}} = -0.45$. Value $\tilde{\Omega}_{1,2}(k) = 0$ corresponds to angular frequency $\Omega_{1,2}(k) = \omega_{n,at}^{(b)} - \omega_{n,at}^{(a)}$ for two-level atoms in the lattice.

The principal feature of dispersion curves in Fig.2 is the presence of minimum for the low branch polaritons when $k = 0$. For small values of quasi-momentum $k\ell \ll 1$ from Eq.(11) one can obtain:

$$\Omega_{1,2}(k) \approx \frac{\hbar k^2}{2m_{1,2}}. \quad (13)$$

The dispersion relation (13) for $\Omega_2(k)$ describes the free quasi-particles (polaritons) at the bottom of the lower curve in Fig.2. Such a peculiarity of the low branch polariton dispersion can be used to achieve the BEC state with quasi-momentum $k = 0$, cf. [11].

Now, let us consider the group velocity $v_{1,2}(k) = \frac{\partial \Omega_{1,2}(k)}{\partial k}$ for polaritons. With the help of Eqs. (11) and (12), it is possible to find out:

$$v_{1,2}(k) = \frac{\hbar \sin(k\ell)}{2\ell m_{ph}} \left( 1 + \frac{m_{ph}}{m_{at}} \pm \left(1 - \frac{m_{ph}}{m_{at}}\right) \frac{\delta}{\left(\delta^2 + 4g^2\right)^{1/2}} \right). \quad (14)$$

For $k\ell \ll 1$, the dependence for group velocities $v_{1,2}$ is approximately linear, i.e. $v_{1,2}(k) \simeq \frac{\hbar k}{m_{1,2}}$, see (13). In this limit, the magnitude of the polariton group velocity is restricted by the inequality

$$2\gamma \ell^2 k \leq v_{1,2}(k) \leq 2\alpha \ell^2 k, \quad (15)$$

where the upper bound corresponds to photon-like polaritons and the lower bound describes the atom-like polaritons. The last case can be achieved under the condition $\frac{g^2}{\tilde{\Delta}^2} \ll \frac{m_{ph}}{m_{at}} \ll 1$.

Thus, the manipulation of group velocity for propagating light in the PolC structure can be realized by varying the polaritonic effective masses $m_{1,2}$ (see Eq.(12)) with the help of the atom-field detuning $\tilde{\Delta}$, and "slow" light phenomenon in PolC.

A complete polariton localization occurs within the Brillouin zone: $v_{1,2} = 0$ for $k\ell = n\pi$, $n = 0, \pm 1,...$. We can explore the effects for spatial storing and retrieving of quantum optical information – cf. [19].

## 3. Polariton properties in the temperature domain

In this section, we analyze the temperature dependence for dispersion of atomic polaritons and also the behavior of their group velocities versus the excitation densities. We consider the properties of the two-level atomic system *without inversion* confined in the cavity and interacting with the optical field under thermal equilibrium for finite (non-zero) temperature.

The operator of excitations $N_{ex,\vec{k}}$ with the quasi-momentum $\vec{k}$ can be defined as

$$N_{ex,\vec{k}} = \psi_{\vec{k}}^{\dagger} \psi_{\vec{k}} + \frac{1}{2} \sum_{\vec{q}} \left( b_{\vec{k}+\vec{q}}^{\dagger} b_{\vec{k}+\vec{q}} - a_{\vec{q}}^{\dagger} a_{\vec{q}} \right). \quad (16)$$

It is easy to see that the operator $N_{ex,\vec{k}}$ commutes with the Hamiltonian $H_{\vec{k}}$ defined in Eq.(5). Therefore, the excitation density $\rho_{\vec{k}}$, defined as

$$\rho_{\vec{k}} \equiv \frac{\langle N_{ex,\vec{k}} \rangle}{N_{tot}} = \lambda_{\vec{k}}^2 + \frac{1}{2N_{tot}} \sum_{\vec{q}} \langle b_{\vec{k}+\vec{q}}^{\dagger} b_{\vec{k}+\vec{q}} - a_{\vec{q}}^{\dagger} a_{\vec{q}} \rangle, \quad (17)$$

is the conserving quantity for the considered atom-field coupled system, where $\lambda_{\vec{k}} = \left( \langle \psi_{\vec{k}}^{\dagger} \psi_{\vec{k}} \rangle / N_{tot} \right)^{1/2}$ is a normalized field amplitude with the momentum $\vec{k}$.

The first term in (17) describes a contribution of the optical field to the density of excitations. The second term $S_{\vec{k}} = \frac{1}{N_{tot}} \sum_{\vec{q}} \left\langle b^\dagger_{\vec{k}+\vec{q}} b_{\vec{k}+\vec{q}} - a^\dagger_{\vec{q}} a_{\vec{q}} \right\rangle$ is responsible for population inversion in the atomic system. Physically, $\lambda_{\vec{k}}$ represents the order parameter in the phase transition problem.

The excitation density $\rho_{\vec{k}}$ is more atom-like when

$$\lambda_{\vec{k}}^2 \ll 1 \qquad (18)$$

Relation (18) implies a small average photon number in comparison with the total number of atoms in the PolC, i.e. $\left\langle \psi^\dagger_{\vec{k}} \psi_{\vec{k}} \right\rangle \ll N_{tot}$.

Some conclusions can be formulated for this case.

First, a low density limit is obtained from Eqs.(17) and (18) and looks like

$$\rho_{\vec{k}} \approx -0.5, \quad S_{\vec{k}} \simeq -1. \qquad (19a,b)$$

In this limit, the polaritons (see Eqs.(8) and (9)) can be treated as the boson quasi-particles.

Second, taking into account Eq.(18), the saturation limit for atomic ensemble is achieved for the excitation density $\rho_{\vec{k}} \approx 0$ ($S_{\vec{k}} \simeq 0$).

Third, for $\rho_{\vec{k}} \approx 0.5$ ($S_{\vec{k}} \simeq 1$) the inversion regime occurs in the two-level atomic system coupled with the optical field.

For large values of the field amplitude, e.g. for $\lambda_{\vec{k}} > 1$, the excitation density $\rho_{\vec{k}}$ becomes more photon-like. In this case, the polaritonic model (in Section 2) breaks down. At the same time, nonlinear effects for atom-field interaction become more important. But this case is not under our consideration and we suppose that relation (18) is to be fulfilled further.

To determine the order parameter $\lambda_{\vec{k}}$ for thermal equilibrium, we use a variational (thermodynamic) approach, see e.g. [28]. In the case, the partition function $Z_{\vec{k}}(N_{tot}, T) = Tr(e^{-\beta H'_{\vec{k}}})$ should be explored; $H'_{\vec{k}} = H_{\vec{k}} - \mu N_{ex,\vec{k}}$ being a modified Hamiltonian, $\beta \equiv (k_B T)^{-1}$. The function $Z_{\vec{k}}(N_{tot}, T)$ describes a grand canonical ensemble with the finite (nonzero) chemical potential $\mu$. The thermodynamic approach under consideration is valid for a large total number of atoms in the PolC structure, i.e. for $N_{tot} = NM \gg 1$. However, the number of atoms $N$ for each site is not so large, and we need to have a large number of cavities ($M \gg 1$) for the case.

In the semiclassical limit, neglecting fluctuations of the optical field and also atom-field correlations, one can obtain

$$\tilde{\omega}_{ph} \lambda_{\vec{k}} = \frac{g^2 \lambda_{\vec{k}} \tanh\left(\frac{\hbar \beta}{2}\left(\tilde{\omega}_{at}^2 + 4g^2 \lambda_{\vec{k}}^2\right)^{1/2}\right)}{\left(\tilde{\omega}_{at}^2 + 4g^2 \lambda_{\vec{k}}^2\right)^{1/2}}, \qquad (20)$$

where we made denotations $\tilde{\omega}_{ph} \equiv \omega_{ph}(k) - \mu$, $\tilde{\omega}_{at} \equiv \omega_{at}(k) - \mu$ and used the coherent basis representation for the order parameter (field amplitude) $\lambda_{\vec{k}}$.

The Eq.(20) is similar to the BCS gap equation applied to study the second-order phase transition to the superfluid (or condensed) state for various systems in solid-state physics, see e.g. [22,28-30].

In quantum optics, this type of phase transition has been discussed many times in the framework of transition to some "superradiant" (coherent or condensed) photonic when the chemical potential $\mu = 0$, see e.g. [28, 33-36]. In fact, the Dicke model and its generalizations (see e.g. [32-36]) for atom-light interaction have been proposed for that. However, in quantum optics, two important circumstances are to be taken into account for the phenomenon.

First, it has been shown in [32] that the main physical feature of the "superradiant" phase transition is connected with building up a spontaneous static field (the field with zero frequency) in the medium, i.e. some ferroelectric state arises in atomic system. But the effect is not related to quantum storage and memory purposes.

Second, it has been pointed out (see e.g. [31]) that the so-called Thomas-Reiche-Kuhn sum rule formally contradicts to nontrivial solution of "superradiant" phase transition for a simple Dicke model (a more recent discussion on the problem is presented in [35]).

Nevertheless Eq.(20) is suitable for the description of phase transition in polaritonic system. In fact, there is a principal peculiarity for the polaritonic system because we need to consider matter-field coupled states even for a simple Dicke model, and then a chemical potential $\mu$ is principally non-zero, cf. [30]. Fixing the total number of excitations $N_{ex,\vec{k}}$, it is possible to obtain a non-zero parameter representing a chemical potential for the case; the parameter is not formally limited by the sum rule mentioned above. Then the photonic condensate ("superradiant" state) should be considered as a limiting (photon-like) state for low branch polaritons, being achieved by choosing an appropriate value for the atom-field detuning $\delta$ such as $\vartheta_1 \approx 0$, $\vartheta_2 \approx 1$, see Eqs.(8) and (9).

Thus, we examine Eq.(20) in the framework of the atomic polaritons phase transition from the normal state to the superfluid (condensed) state. The parameter combination $2g\lambda_{\vec{k}}$ plays a role of the frequency gap, as the normalized photon number $\lambda_{\vec{k}}$ represents the order parameter. The normal state of polaritons is obtained from Eq.(20) for $\lambda_{\vec{k}} = 0$, and occurs for temperature $T \geq T_c$ ($T_c$ is the critical temperature of the phase transition). On the other hand, there is a nontrivial solution of Eq.(20) with $\lambda_{\vec{k}} \neq 0$ characterizing the superfluid state of polaritons for the temperature $T < T_c$.

Two average collective parameters give an important feature for a two-level atomic system without inversion, i.e. the stationary polarization $P_{\vec{k}}(t) = P_{\vec{k}} e^{-i\mu t}$ and the population imbalance for the atomic levels $S_{\vec{k}}$, that looks like:

$$P_{\vec{k}} = -\frac{g\lambda_{\vec{k}}\tilde{\omega}_{at}\tanh\left(\frac{\hbar\beta}{2}\left(\tilde{\omega}_{at}^2 + 4g^2\lambda_{\vec{k}}^2\right)^{1/2}\right)}{|\tilde{\omega}_{at}|\left(\tilde{\omega}_{at}^2 + 4g^2\lambda_{\vec{k}}^2\right)^{1/2}}, \tag{21a}$$

$$S_{\vec{k}} = -\frac{|\tilde{\omega}_{at}|\tanh\left(\frac{\hbar\beta}{2}\left(\tilde{\omega}_{at}^2 + 4g^2\lambda_{\vec{k}}^2\right)^{1/2}\right)}{\left(\tilde{\omega}_{at}^2 + 4g^2\lambda_{\vec{k}}^2\right)^{1/2}}, \tag{21b}$$

For the normal state of polaritons ($\lambda_{\vec{k}} = 0$) for $T \geq T_c$ from Eq.(21) we obtain

$$P_{\vec{k}} = 0, \quad S_{\vec{k}} = -\tanh\left(\frac{\hbar|\tilde{\omega}_{at}|}{2k_B T}\right). \tag{22a,b}$$

Availability of the solution of Eq. (20) with $\lambda_{\vec{k}} \neq 0$ for two-level atoms in the cavity array (condensed state) results in stationary polarization $P_{\vec{k}}(t)$ appearing on *non-zero frequency*, which is equal to the chemical potential $\mu \neq 0$. The angular frequencies $\tilde{\omega}_{ph}$ and $\tilde{\omega}_{at}$ in Eq.(20) determine the detunings of both the optical field frequency $\omega_{ph}(k)$ and the frequency of two-level atomic transition $\omega_{at}(k)$ from the frequency $\mu$ of polarization in atomic ensemble.

By using Eq.(21) and the partition function $Z_{\vec{k}}(M,T)$, we can get an expression for the polariton excitation density $\rho_{\vec{k}}$ versus temperature

$$\rho_{\vec{k}} = \lambda_{\vec{k}}^2 - \frac{1}{2}\frac{\left|\tilde{\omega}_{at}\right|\tanh\left(\frac{\hbar\beta}{2}\left(\tilde{\omega}_{at}^2 + 4g^2\lambda_{\vec{k}}^2\right)^{1/2}\right)}{\left(\tilde{\omega}_{at}^2 + 4g^2\lambda_{\vec{k}}^2\right)^{1/2}}. \tag{23}$$

The phase transition to the superfluid/condensed state occurs for the critical excitation density $\rho_{\vec{k},c} = -\frac{1}{2}\tanh\left(\frac{\hbar\left|\tilde{\omega}_{at}\right|}{2k_B T_c}\right)$ obtained from Eq.(23) for the critical temperature $T = T_c$ of the phase transition.

Putting Eq.(23) into Eq.(20) we obtain for the chemical potential $\mu$ of the polaritonic system:

$$\mu_{1,2} = \frac{1}{2}\left[\omega_{at}(k) + \omega_{ph}(k) \pm \omega_{R,eff}(k)\right], \tag{24}$$

where $\omega_{R,eff}(k) = \left(\delta^2 + 4g_{eff}^2\right)^{1/2}$ is the effective Rabi splitting angular frequency; $g_{eff} = g\left(2\left(\lambda_{\vec{k}}^2 - \rho_{\vec{k}}\right)\right)^{1/2}$ defines the effective atom-field coupling coefficient depending on the temperature $T$, excitation density $\rho_{\vec{k}}$ and normalized field amplitude $\lambda_{\vec{k}}$. In the atom-field interaction for the resonance condition $\delta = 0$ and $\rho_{\vec{k}} = \lambda_{\vec{k}}^2$, we have for Rabi splitting frequency $\omega_{R,eff}(k) = 0$. A positive density $\rho_{\vec{k}} > 0$ corresponds to inversion in the two-level atomic system.

We now focus on the polariton properties of the atomic system without inversion, i.e. when the excitation density is negative ($\rho_{\vec{k}} < 0$). In this limit, the Eqs.(24) defines the upper ($\mu_1$) and lower ($\mu_2$) branches for polariton frequencies versus both temperature and excitation density. From Eqs.(24) and using expression (19) we arrive at expression (11) for low density limit. In the inset of Fig.2, the red curves demonstrate polariton dispersion for the non-zero order parameter $\lambda_{\vec{k}}$ and for the excitation value $\rho_{\vec{k}} = -0.45$.

By using (24) we can bring Eq.(20) to the form:

$$\Omega_{ph} = \frac{\tanh\left(\frac{\hbar\beta g}{2}\left(\Omega_{at}^2 + 4\lambda_{\vec{k}}^2\right)^{1/2}\right)}{\left(\Omega_{at}^2 + 4\lambda_{\vec{k}}^2\right)^{1/2}}, \tag{25}$$

where we made denotations: $\Omega_{at} = -\delta/2g + \left(\left(\delta/2g\right)^2 + 2\left(\lambda_{\vec{k}}^2 - \rho_{\vec{k}}\right)\right)^{1/2}$,

$\Omega_{ph} = \delta/2g + \left(\left(\delta/2g\right)^2 + 2\left(\lambda_{\vec{k}}^2 - \rho_{\vec{k}}\right)\right)^{1/2}$.

The critical temperature $T_c$ of the phase transition between normal and superfluid (BEC) states can be easily derived out from Eq.(25) by substituting $\lambda_{\vec{k}} = 0$. One can obtain:

$$T_c \simeq \frac{\hbar g \Omega_{at}^{(0)}}{2k_B \text{at anh}\left(-2\rho_{\vec{k}}\right)} \approx \frac{\hbar g \left(-2\rho_{\vec{k}}\right)^{1/2}}{k_B \ln\left(2(1+2\rho_{\vec{k}})^{-1}\right)}, \quad (26)$$

where $\Omega_{at}^{(0)} \equiv \Omega_{at}\big|_{\lambda_{\vec{k}}=0}$. The last expression in Eq.(26) is true for $\delta = 0$ in the low-temperature limit, when the condition

$$k_B T \ll \hbar g \quad (27)$$

is satisfied.

In Fig.3, the phase diagram (the phase boundary) for $T_c$ according to Eq.(26) is presented. It is clearly seen that a very low excitation density requires a large value of the coupling constant $g$ for high enough temperature of the phase transition. On the other hand, for a given value of the $g$ parameter, the phase transition temperature $T_c$ decreases when the density of excitations approaches the value corresponding to the low density limit described by condition (19). The critical temperature becomes zero for positive detuning, see Fig.3b. The fact is connected with "heavy" (i.e. atom-like) polaritons for the lower branch, see Eq.(12). On the other hand for negative detuning ($\delta < 0$) the polaritons become more photon-like, and the temperature $T_c$ increases.

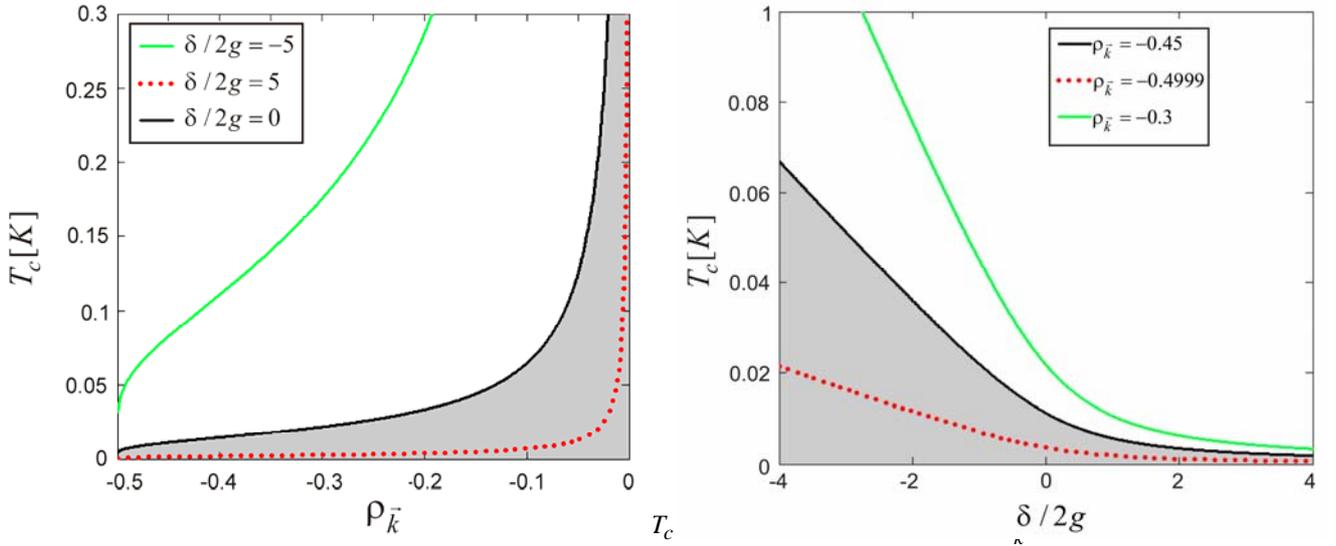

filed frequency detuning $\delta/2g$. The shaded region corresponds to the superfluid state of polaritons for (a) $\delta = 0$ and (b) $\rho_{\vec{k}} = -0.45$.

In Fig.4, we plotted the dependences for the order parameter $\lambda_{\vec{k}}$ as a function of $T/T_c$ in accordance with numerical solution of Eq. (25) for various detunings $\delta$. The dependences in Fig.4 demonstrate a second-order continuous phase transition for the photon number in the cavity array that occurs for low enough (*mK*) temperature. The zero temperature order parameter $\lambda_{\vec{k},0} \equiv \lambda_{\vec{k}}\big|_{T=0}$ decreases for the positive detuning $\delta$ when the low branch polaritons become more atom-like. In this region for the order parameter we have

$$\lambda_{\vec{k},0} \simeq \left(\frac{1-4\rho_{\vec{k}}^2}{8\Omega_{ph}^{(0)}\left((\delta/2g)^2 - 2\rho_{\vec{k}}\right)^{1/2}}\right)^{1/2}, \quad (28)$$

where $\Omega_{ph}^{(0)} \equiv \Omega_{ph}\big|_{\lambda_{\vec{k}}=0}$.

In the low temperature limit (27), the behavior of the order parameter, presented in Fig.4, is described by the expression:

$$\lambda_{\vec{k}} \simeq \lambda_{\vec{k},0}\left(1-\exp\{x_c - x\}\right)^{1/2}, \tag{29}$$

where we introduce the dimension-less parameter $x = \hbar g \beta \left|\Omega_{at}^{(0)}\right|$, $x_c = x\big|_{T=T_c}$.

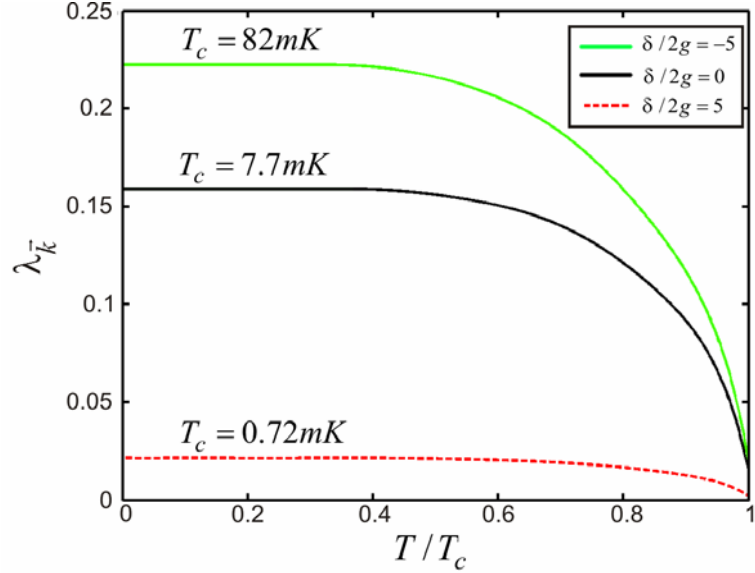

**Figure 4**. Dependence for the order parameter $\lambda_{\vec{k}}$ versus reduced temperature $T/T_c$. The excitation density $\rho_{\vec{k}} = -0.45$.

For high enough temperatures when $k_B T \gg \hbar g$, and for $T \to T_c$ from Eq.(25) under condition (18) for the order parameter $\lambda_{\vec{k}}$ we have:

$$\lambda_{\vec{k}} \simeq \left(\frac{\left(\Omega_{at}^{(0)}\right)^2 \Omega_{ph}^{(0)} x_c}{8\sinh(x_c)\left((\delta/2g)^2 - 2\rho_{\vec{k}}\right)^{1/2}}\left(1-\frac{T}{T_c}\right)\right)^{1/2}. \tag{30}$$

Experimentally, it is easier to study the dependence for the order parameter $\lambda_{\vec{k}}$ versus the atom-field detuning $\delta$ or versus angular frequency of the optical field $\omega_{ph}(k)$ when the temperature of the atomic system is fixed. Near the critical point, the dependences in Fig.4 can be described as [cf.(30)]

$$\lambda_{\vec{k}} \simeq \left(\frac{\hbar\omega_{ph,c}\left(\Omega_{at,c}\right)^3 \Omega_{ph,c}}{16 k_B T \sinh(x_c)\left[(\delta_c/2g)^2 - 2\rho_{\vec{k}}\right]}\left(1-\frac{\omega_{ph}}{\omega_{ph,c}}\right)\right)^{1/2}, \tag{31}$$

where we denote $x_c = \hbar g \beta |\Omega_{at,c}|$ and $\Omega_{at,c} = -\delta_c/2g + \left((\delta_c/2g)^2 - 2\rho_{\tilde{k}}\right)^{1/2}$, $\Omega_{ph,c} = \delta_c/2g + \left((\delta_c/2g)^2 - 2\rho_{\tilde{k}}\right)^{1/2}$; $\delta_c = \omega_{ph,c} - \omega_{at}$ is a critical detuning, $\omega_{ph,c}$ is a critical angular frequency of optical field for which the phase transition to the state with $\lambda_{\tilde{k}} = 0$ occurs. Such a critical frequency $\omega_{ph,c}$ depends on the field distribution in a spatially periodic system (see Fig.1).

The order parameter dependence on temperature results in the temperature dependence of frequency gap between two branches (see Fig.2). The dependence for the resonant effective Rabi splitting frequency $\omega_{R,eff}/2\pi$ is presented as a function of the normalized temperature $T/T_c$ in Fig.5. The frequency gap is minimal, i.e. $\omega_{R,norm}(k) = \left(\delta^2 - 8g^2\rho_{\tilde{k}}\right)^{1/2}$, for the normal state of polaritons. For relatively small photon number in the cavity array, i.e. under condition (18) for superfluid polaritons, the effective Rabi splitting angular frequency $\omega_{R,eff}(k)$ can be represented as

$$\omega_{R,eff}(k) \approx \omega_{R,norm}(k) + \frac{4g^2 \lambda_{\tilde{k}}^2}{\left(\delta^2 - 8g^2\rho_{\tilde{k}}\right)^{1/2}}, \qquad (32)$$

The last term in Eq.(32) describes the additional frequency gap between polaritonic branches; the temperature dependence occurs due to the phase transition for $T < T_c$, see Fig.5.

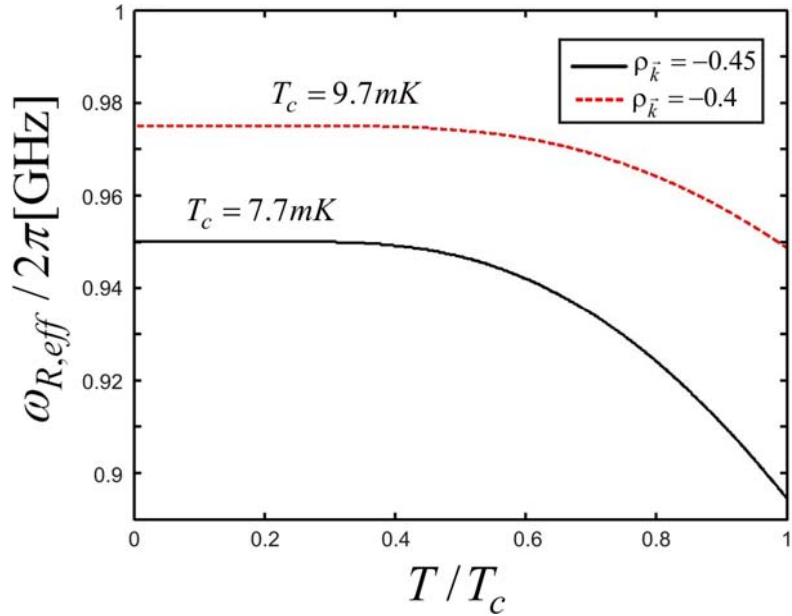

**Fig.5**. Dependence of the resonant ($\delta = 0$) effective Rabi splitting frequency on the reduced temperature $T/T_c$. The value $\omega_R/2\pi = g/\pi = 1GHz$ corresponds to normal Rabi splitting frequency for rubidium atoms interacting with the photonic field.

Now let us examine the polariton group velocity behavior versus temperature when the equilibrium state of the system occurs. For the case we can use Eqs.(12)-(14) taking into consideration the effective atom-light coupling parameter $g_{eff}$ instead of the parameter $g$.

In Fig.6, we present the phase boundaries for the group velocities of low branch polaritons versus reduced temperature $T/T_c$ of the atomic system. The parameters chosen in Fig.6 correspond to atom-like polaritons with small $k$ in the bottom region of the dispersion curve (see Fig.2); polaritons having approximately the same critical temperature under the condition $|\tilde{\delta}| \gg |\alpha - \gamma| k^2 \ell^2$.

In limit $\tilde{\Delta} \gg 2|g_{eff}|$ and $\frac{m_{ph}}{m_{at}} = \frac{\gamma}{\alpha} \ll \frac{g_{eff}^2}{\tilde{\Delta}^2} \ll 1$ we can obtain for the polariton group velocity

$$v_2(k) \approx \frac{2\hbar k g^2 \left(\lambda_{\vec{k}}^2 - \rho_{\vec{k}}\right)}{m_{ph}\tilde{\Delta}^2} = \frac{4\alpha \ell^2 k g^2 \left(\lambda_{\vec{k}}^2 - \rho_{\vec{k}}\right)}{\tilde{\Delta}^2}. \qquad (33)$$

Equati.(33) demonstrates a significant reduction of the group velocity value for atom-like polaritons. In fact, we obtain $v_2(k)/c \simeq 10^{-5}$ under conditions of Fig.6. The expression $v_{2,c}(k) = \frac{4\alpha \ell^2 k g^2 |\rho_{\vec{k},c}|}{\tilde{\Delta}^2}$ describes a critical velocity of low branch polaritons (for given $k$) (ses Eq.(31) for $T = T_c$). On the other hand, the maximal group velocity of polaritons $v_{2,0}(k) = \frac{4\alpha \ell^2 k g^2 |\rho_{\vec{k}}|}{\tilde{\Delta}^2}$ is achieved when the order parameter $\lambda_{\vec{k}} = \lambda_{\vec{k},0}$ for $T = 0$.

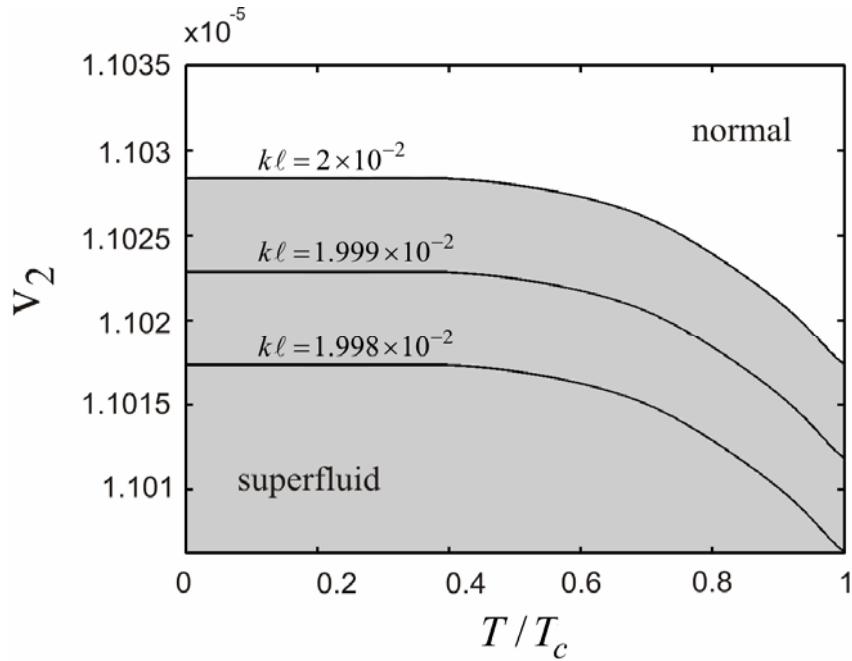

**Fig.6.** The reduced group velocity $v_2 \equiv v_2(k)$ for the lower branch polaritons as a function of reduced temperature $T/T_c$ (the detuning $\tilde{\delta}/2g = 5$); $v_2$ is represented in units of the light velocity in vacuum. The shaded region corresponds to the superfluid state for atom-like polaritons with various quasi-momentum $k$.

There exist some temperature domain and the order parameter value region (see Fig.4) where the polaritons (for given quasi-momentum $k$) fall in the superfluid state. In the case for the velocity value we have $v_{2,c}(k) \leq v_2(k) < v_{2,B}(k)$, where $v_{2,B}(k)$ is a velocity of polaritons in the phase boundary of transition to superfluidity.

## 4. Conclusion

We have developed the theory of formation of coherent polaritons in a spatially periodic structure for trapped two-level atomic ensembles. A quantum analysis has been carried out for the interaction of atoms with optical field in the cavity arrays under the tight-binding approximation. We have shown that the macroscopic polarization of the atomic medium at non-zero frequency occurs in the case. We have obtained the expression for a many-body Hamiltonian that describes the atom-field interaction in momentum space. The approach has been applied for some problems of phase transition from the normal state to the superfluid state for low branch polaritons. We have shown that the observable group velocity for propagating optical wave packet can be essentially reduced due to the localization of atom-field excitations (polaritons) in the polaritonic crystal structure. Using a thermodynamic approach, we have considered the problem of BCS state formation for polaritons in the one-dimensional atom-optical array. We mainly focused our analysis on the low excitation density approximation for which the average number of photons is much smaller than the number of atoms. We have found out that the second order continuous phase transition results in formation of the superfluid state of polaritons with some certain behavior of group velocity. The fact can be useful to determine the superfluid properties of polaritons in the medium.

**Acknowledgments**

This work is partially supported by Russian Foundation for Basic Research (grants No 08-02-99011, 09-02-91350) and by some other grants from Agency of Education and Science of Russian Federation. One of us (A.P.A.) is grateful to participants 450.WE-Heraeus-Seminar "Mixed States of light and Matter" for fruitful discussions. In particular, we acknowledge stimulating discussions with Professors Martin Weitz, Peter Littlewood and Michael Fleischhauer.